# A promising candidate for ising ferromagnetism of two-dimensional kagome $V_2O_3$ honeycomb monolayer

Fazle Subhan [1], Chuanhao Gao[1], Luqman Ali[2], Yanguang Zhou[*,3], Zhenzhen Qin[*,4], and Guangzhao Qin[*,1,5,6]

| | |
|---|---|
| 1 | State Key Laboratory of Advanced Design and Manufacturing Technology for Vehicle, College of Mechanical and Vehicle Engineering, Hunan University, Changsha 410082, P. R. China |
| 2 | Department of Pharmacology, University of Virginia, 22903, Charlottesville, VA, USA |
| 3 | Department of Mechanical and Aerospace Engineering, The Hong Kong University of Science and Technology, Clear Water Bay, Kowloon, Hong Kong SAR |
| 4 | International Laboratory for Quantum Functional Materials of Henan, and School of Physics and Microelectronics, Zhengzhou University, Zhengzhou 450001, P. R. China |
| 5 | Research Institute of Hunan University in Chongqing, Chongqing 401133, China |
| 6 | Greater Bay Area Institute for Innovation, Hunan University, Guangzhou 511300, Guangdong Province, China |

**Abstract:** Due to the low dimensionality in the quantization of the electronic states and degree of freedom for device modulation, two-dimensional (2D) ferromagnetism plays a critical role in lots of fields. In this study, we perform first-principles calculation to investigate the ising ferromagnetism and half-metallicity of kagome $V_2O_3$ monolayer. Based on the calculations using different functional, it is found that GGA-PBE gives a half-metallic band gap while the GGA+U gives a semiconductor narrow band gap (~1.1 meV), which shows quasi-half metallic nature. By studying the magnetic properties with LDA, GGA-PBE, and GGA+U, we get a robust ferromagnetic ground state, where the giant perpendicular magnetic anisotropy energy of ~0.544 meV is achieved by applying the spin-orbit coupling (SOC) with GGA+U. Furthermore, by exploring the orbital contribution to the electronic bands and the magnetic crystalline anisotropy, it is uncovered that the 3d (V) orbitals contribute to the *out-of-plane*. The electronic band structure shows two flat bands ($F_1$ and $F_2$) and Dirac points ($D_1$ and $D_2$) which further confirm that kagome $V_2O_3$ ML can also be used for topological properties. Besides, the Curie temperature of the $V_2O_3$ ML is calculated to be 640 K by Metropolis Monte Carlo (MC) simulations.

Correspondence: maeygzhou@ust.hk  (Y. Z), qzz@zzu.edu.cn  (Z. Q), gzqin@hnu.edu.cn  (G. Q);

## 1. Introduction

The successful isolation of single layers of carbon atoms in form of first two dimensional (2D) material called graphene[1] results in promising device applications. However, due to some limitations in the intrinsic properties of graphene, scientists have been motivated to explore many other 2D materials including hexagonal boron nitride, transition metal dichalcogenides, phosphorene, *etc*, which are synthesized and exfoliated from their bulk with different experimental techniques and tremendously studied both experimentally and theoretically.[2–5] Due to the low thermal conductivity and high Figure of merit, 2D materials were extensively studied for the thermal transport properties[6–12]. Furthermore, the literature shows that, the study of the nanoribbons for the electronic transport properties also confirms the intrinsically non-magnetic nature of the 2D materials[13–15]. As most of these materials are intrinsically non-magnetic. Therefore, one of the primary goal in 2D materials is to develop ferromagnetic (FM) semiconductors, which not only are eagerly needed in the next generation nano-spintronics, but also exhibit unusual magnetism which offer a great interest in this field.[16,17] However, for the spintronic applications, the lack of robust intrinsic ferromagnetic behavior lock the development of the next generation ferromagnetic 2D materials.[18] Although, the earlier reports shows that the magnetic ground state can be induced in these non-magnetic materials by some external mediations; such as substitutional doping, impurity adsorption, or vacancy defect [19,20]. Due to the external mediation, there exist no long-range magnetism in the materials. Thus, for spintronics applications, a highly desirable ferromagnetic is more favorable and the search for such materials was the focus of the research in the last decade.

Interestingly, in the recent past, some 2D materials has been extensively studied both experimentally and theoretically, called transition metal tri-iodide ($XI_3$) such as $CrI_3$ and $VI_3$ from the bulk down to few and monolayer [21–23]. These materials have intrinsic ferromagnetic ground state with giant perpendicular magnetocrystalline anisotropy. Besides, a *van der Waal* 2D material called $Cr_2Ge_2Te_6$ (CGT) has been extensively studied from monolayer to few layer with low perpendicular magnetocrystalline anisotropy and Curie temperature with an indirect band gap.[24–26] H.J Deiseroth *et al.* also reported the FM state in $Fe_3GeTe_2$ (FGT) [27]. It was reported in FGT an intrinsic ferromagnetic nature with a high Curie temperature of ~200K with metallic nature and perpendicular anisotropy in the monolayer.

Spintronics is a prominent filed in condensed matter physics, which provides the utilization of spin degree of freedom in various practical applications for various innovative technologies[28]. Interestingly, for spintronic applications, half metallic materials play a vital role. Because the coexistence of one spin in metallic and other in semiconductor nature and avails a prominent role in the spin-polarized current generation and injection[29,30]. Although, numerous studies of half metallic materials were in the bulk forms including Heusler alloys[31], double perovskites[32], transition metals oxides and chalcogenides[33–35], while few are in the 2D materials[36].

The current developments in two-dimensional material intriguer our attention towards the study of the oxides materials. Besides, the versatility of these materials makes them attractive for the spintronics applications. Interestingly, these observations of Dirac points and flat bands in the vicinity of $E_F$ may attract the researchers for manipulating the topological properties in magnetic or half-metallic materials[37]. The thoroughly investigated different behavior of the $V_2O_3$ kagome monolayer with different ab-initio potentials would shed light on future studies on ferromagnetism and next generation spintronic applications. In this study, we thoroughly investigated of $V_2O_3$ ML using different first principles methods and found some interesting results as reported below.

2. **Numerical Methods**

**2.1** First principles calculations details

We use the density functional theory (DFT) calculations with projected augmented wave (PAW) under the Vienna *ab initio* simulation package (VASP)[38,39]. The generalized gradient approximation (GGA) was adopted in Perdew-Burke-Ernzerhof (PBE)[40]. For the plane wave cut-off energy of 500 eV was used. The Brillouin zone has been sampled by the Monkharst-Pack method [41] within atomically generated k-mesh of 11x11x1. The convergence criterion for energy was set to 0.01 meV, while the atomic positions were fully relaxed until the Hellmann-Feynman force on each atom becomes smaller than $10^{-4}$ eV/Å. To avoid the artificial interactions between the images more than 20 Å vacuum along the z-axis was applied.

**2.2.** Phonon and ab-initio molecular dynamics

For phonon stability, we use the phonopy code [42] with a force criterion of $10^{-8}$ eVÅ$^{-1}$. Therefore, for phonon calculations, we use a supercell of 2x2x1 with a k-mesh of 13x13x1. Further, to check the thermal stability of the kagome $V_2O_3$ ML, ab initio molecular dynamics simulations (AIMD) method were performed at room temperature with temperature step of 5ps.

**2.3.** Magnetic anisotropy calculation:

For magnetic anisotropy energy (MAE) calculations, we use the non-collinear energy calculations including spin orbit coupling (SOC). The energy criterion is set up to $10^{-8}$ with a denser k-mesh of 21x21x1. Further, we use different functional as given in Table 1. For, MAE we use two magnetic axes [1 0 0] or [0 1 0], and [0 0 1], which are respectively known as in-plane or parallel ($E_{in}$) and out of plane or perpendicular ($E_{out}$) directions. The final MAE is calculated from using the following equation MAE

$$MAE = E_{in} - E_{out} \qquad (1)$$

## 3. Results and discussions

### 3.1 Structural Stability

First, we explored the phonon and thermal stability of the kagome $V_2O_3$ ML. From the phonon dispersion, we found no imaginary modes, which confirms the dynamically stability of kagome $V_2O_3$ ML. Fig. 1 (a-b), shows the phonon band structure and projected density of states respectively, where the V bands are strongly localized and they occupied the low frequency modes below 5 THz, while the high frequency modes in the band structure is mainly occupied by the O atoms. Above the 20 THz frequency, O bands are dominated. Further, for thermal stability we use the ab-initio molecular dynamic (AIMD) at room temperature. Figure 1 (c-d) shows the top and side views of the ML at the end of 5 ps at 300 K. These investigations confirm that $V_2O_3$ ML is experimentally feasible to be fabricated.

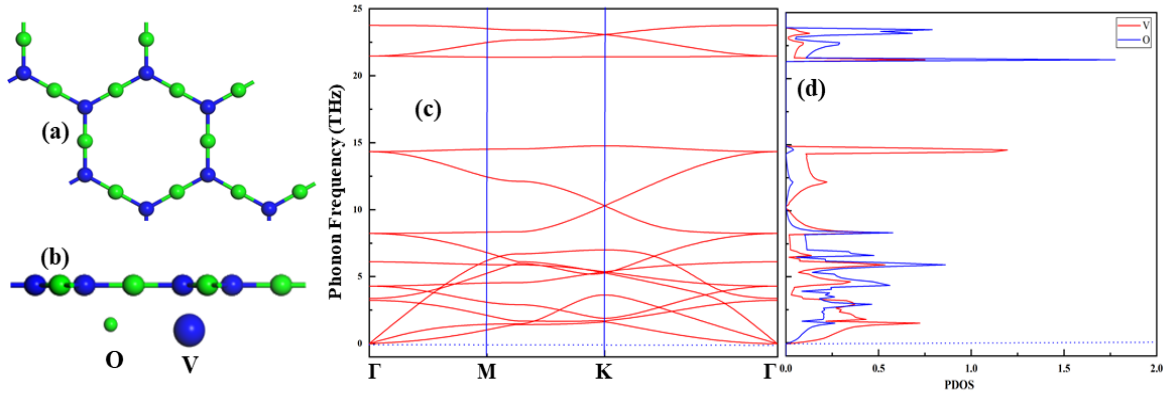

**Figure.1** Thermal and dynamic stability, top (a) and (b) side views of $V_2O_3$ ML at the end of a 5-ps ab initio MD simulation at 300 K. Phonon (c) band structure, (d) projected DOS of kagome $V_2O_3$

Figure 1 (a) shows the crystal structure of two dimensional (2D) $V_2O_3$ ML, where each V atom is surrounded by three O atoms in a hexagonal form, and the O atoms form the kagome lattice. The bond length of V-O and bond angle O-V-O before (after) the structural relaxation is 1.804 (1.757 Å) and 120.1° (120.08°) respectively. Further, we also calculated the formation energy using the following equation

$$E_f = \left[ \frac{E(V_2O_3) - 2E(V) - 3\mu(O_2)}{5} \right] \qquad (2)$$

Where $E_f$ is the formation energy per atom of the kagome $V_2O_3$ ML. E ($V_2O_3$) is the total energy of the $V_2O_3$ monolayer (ML), E(V) is the energy of single V atom in the stable bulk (BCC) is -9.32 eV while $\mu O_2$ is the chemical potential of the single O gas is -2.60 eV respectively. Interestingly, we found that the formation energy of the $V_2O_3$ ML per atom is -3.02 eV/atom, which is not only matching well with the previous reported systems [43,44] but also with the $V_2O_3$ bulk (-2.29 eV/atom).

## 3.2 Magnetic Properties

To find the energetically stable magnetic ground state of the kagome $V_2O_3$ monolayer, we calculated the total energies for both the ferromagnetic (FM) and antiferromagnetic (AFM) configurations of the system. In case of AFM, we consider three different configurations as shown in Figure 2 (a-d), and calculate the total energy for each configuration. The energy difference between FM and AFM was calculated by using the following equation.

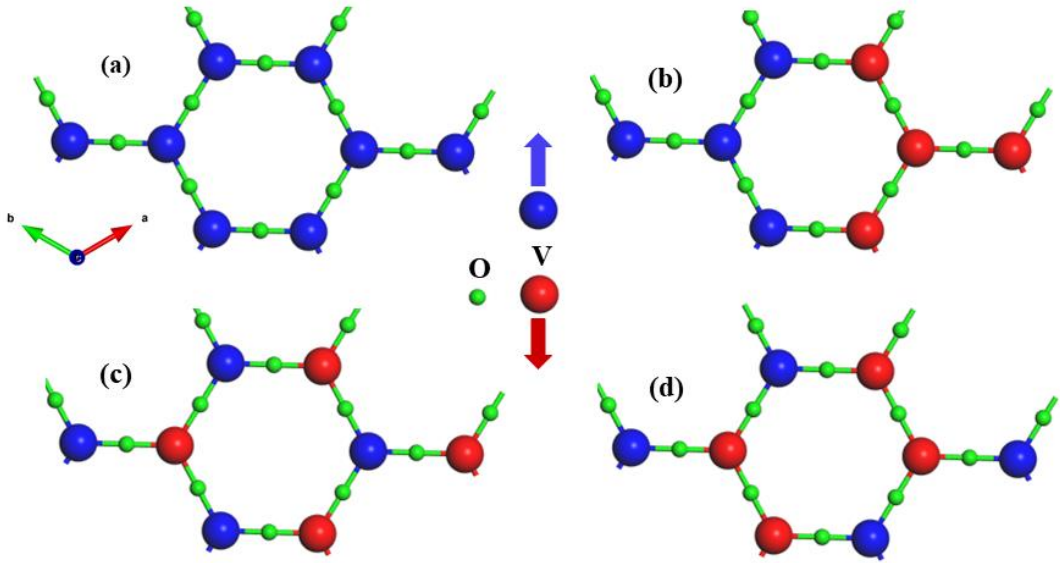

**Figure 2.** Structural illustration of $V_2O_3$ monolayer for magnetization of (a) FM, (b) $AFM_1$, (c) $AFM_2$, (d) $AFM_3$. The blue (red) and green balls shows the spin up (spin down) for V atoms and O atom respectively.

The energy difference between FM and AFM was calculated by using the following equation.

$$\Delta E = E_{FM} - E_{AFM} \quad (3)$$

Where $\Delta E$, $E_{FM}$ and $E_{AFM}$ shows the energies difference between ferromagnetic ($E_{FM}$) and antiferromagnetic ($E_{AFM}$) ground state respectively. Using the above relation, we found that $AFM_2$ is energetically more stable configuration as compare to other configurations irrespective of the functional. Further, the magnetic moment remains unchanged irrespective of the functional used for calculation as given in Table 1. According to Hund's rule each electron of $V_2O_3$ ML occupying the $t_{2g}$ triplet state of $V_2O_3$ ML with $S = \frac{3}{2}$. We also calculated the energy differences between the FM and AFM configurations using the different DFT functional and reported in Table 1 below. Further, the magnetic moment per unit cell and per V atom is also given in this table. Interestingly, the magnetic moment remains unchanged irrespective of the functional used in our calculations.

**Table 1**: The energy difference $\Delta E = E_{FM} - E_{AFM}$ using different functional. Magnetic moment per V atom and per cell $\Delta E$ measure in meV while the magnetic moment in $\mu_B$ respectively.

| Functional | ΔE (meV) | Magnetic moment /V atom | Magnetic moment/cell |
|---|---|---|---|
| **LDA** | 299.65 | 2 | 4 |
| **GGA-PBE** | 324.57 | 2 | 4 |
| **GGA+U** | 409.87 | 2 | 4 |

Next, we need to calculate the magnetic anisotropy energy of the kagome $V_2O_3$ ML. For this we performed non-collinear calculations by including the spin orbit coupling (SOC) at a denser k-mesh of 17x17x1. We carefully check the energy convergence calculations. As previous reports shows that perpendicular magnetic anisotropy (out of plane) is playing a key role in the development of magnetic random memory devices due its rich physical phenomenon [45,46]. The fascinating physics of spin Hall switching along with the skyrmions is because of this property in nanostructures like materials [47–49]. Therefore, we consider the magnetic anisotropy. In this regard we consider two magnetic axes along [001] and [100] directions called the out of plane and in plane magnetic anisotropy because the other direction [010] is an analogue of [100]. Interestingly, in case of $V_2O_3$ ML, including spin-orbit interactions, after the total energy convergence calculations we found that the magneto-crystalline anisotropy is in the perpendicular direction having a value of 0.544 meV.

Now to discuss an intuitive picture of MAEs, as we know that the MAE is basically comes from the spin-orbit interaction. Basically SOC is very sensitive to the orbitals of both the occupied and unoccupied spin states. Therefore, the following equation is a confirmation of the above statement.

$$MAE = \xi^2 \sum_{u,o;\alpha,\beta}(2\delta_{\alpha\beta} - 1)\left[\frac{|<u,\alpha|\hat{L}z|o,\beta>|^2 - |<u,\alpha|\hat{L}x|o,\beta>|^2}{\varepsilon_{u,\alpha} - \varepsilon_{o,\beta}}\right] \quad (4)$$

In this equation, $\xi$ is the strength of spin-orbit coupling and $\varepsilon_{u,\alpha}$, $\varepsilon_{o,\beta}$ in the denominator shows the energy levels of unoccupied and occupied states with spin α and β respectively. Further, as for each spin the magnetic moment is related to the electrons of the occupied states, therefore the magnetic moment is insensitive to the orbital character, while the MAE is strongly dependent on the orbital feature, because of the SOC mechanism. Following the second order perturbation theory, we use the SOC which further allows to calculate the matrix elements difference between the easy and hard axis for the angular momentum (l = 1, 2). Thus, we analyzed contribution to the MAE from each SOC channel as shown in Figure 3. The V d and O P orbital contribution is respectively shown in Figure 3 (a)-(b). The positive and negative value means the perpendicular and in-plane contributions to the magnetic anisotropy. For each V atom, the SOC through ($d_{xz}$, $d_z^2$) orbitals has the largest contribution to the perpendicular magnetic anisotropy while the ($d_{yz}$, $d_{xz}$) has also some prominent contribution to the perpendicular magnetic anisotropy. Similarly, the ($d_{xy}$, $d_{xz}$) and ($d_{yz}$, $d_{x^2-y^2}$) has very small contribution in the in-plane magnetic anisotropy as shown in Figure 3 (a).

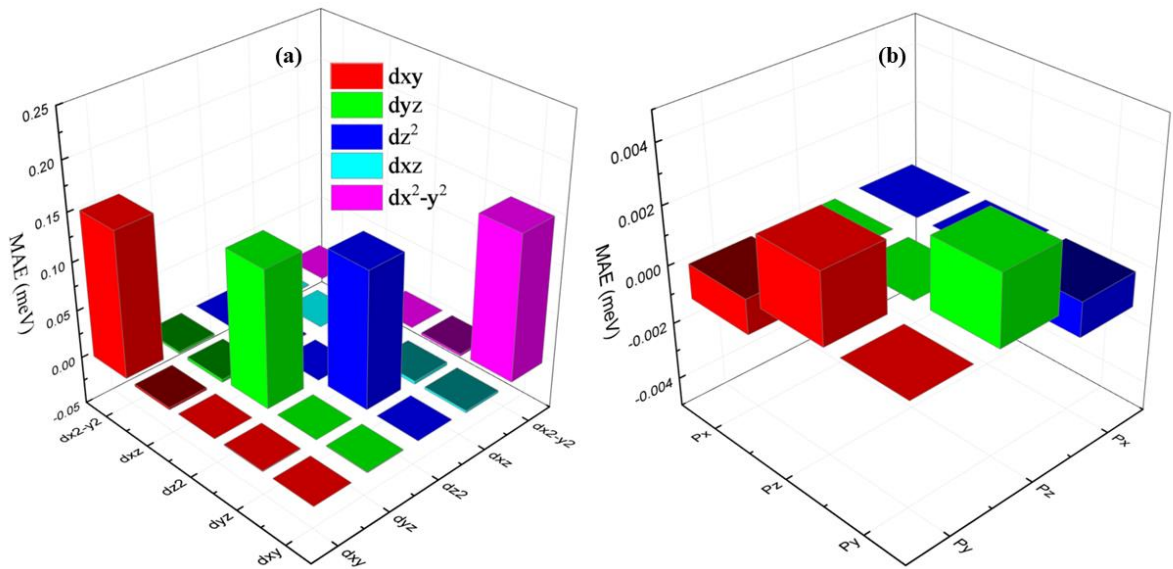

**Figure. 3** Atom resolved MAE of $V_2O_3$ monolayer (a) *d*- resolved MAE of V atoms and, (b) *p*- resolved MAE of O atoms.

Furthermore, each O atom has a negligible negative value between ($P_x$, $P_y$) orbitals results in the in-plane magnetic anisotropy as shown in Figure 3 (b). Therefore, overall the V atoms are responsible for the magnetic anisotropy in the out of plane direction while the O atom contribution is negligible.

Now to discuss the relation between angular dependency and spin orientation of the MAE along polar (ϕ) and azimuthal angles (θ), we rotate the polar angle through the plane of a and b axis of kagome $V_2O_3$ ML while keeping the spin orientation along c axis as shown in Figure 4.

It is noticed that, if we rotate the system at ϕ = 0 to 180° with the spin orientation parallel to the c-axis, we found that MAE is maximum at ϕ = 90° while the increase/decrease of ϕ from 0 to 180° causes decrease in the MAE as shown by the blue dashed lines. Here the spin orientation is along c-axis or (out of plane) of the azimuthal angle, spins are perpendicular to the plane containing a and c axis. Thus we get out of plane MAE. Similarly, if we rotate the spins in the ab plane, the MAE will be remains zero for the same rotation of the polar angle as shown by the red dashed lines in Figure 4.

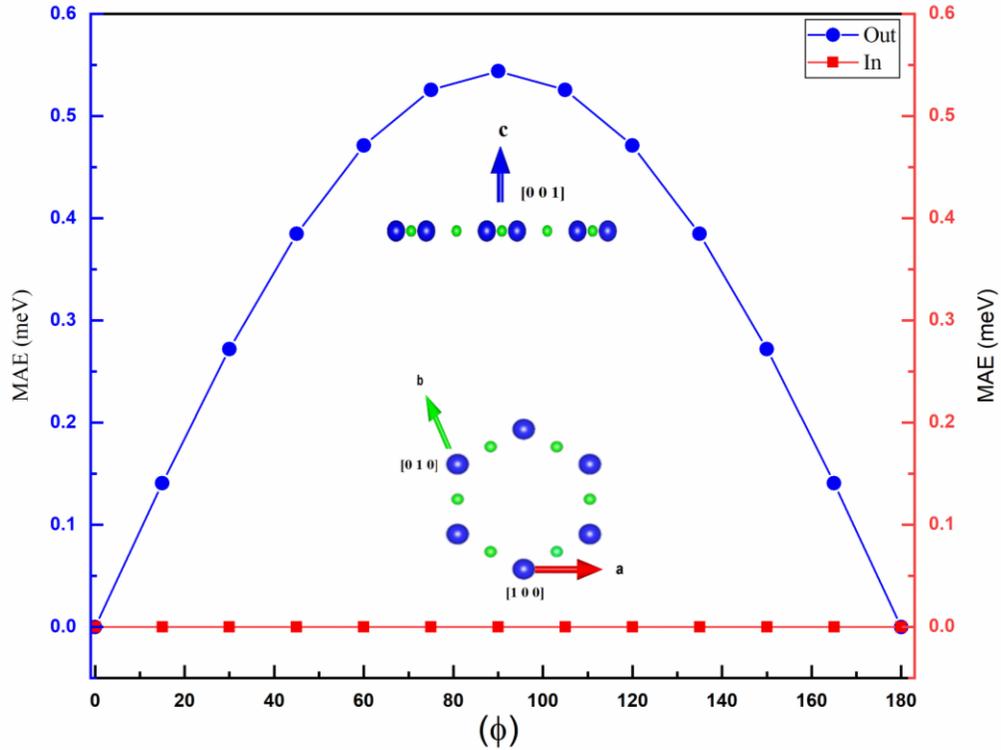

**Figure 4.** Angular dependence of MAE of $V_2O_3$ ML on $\theta$ while $\phi = \pi/2$ as shown by blue dashed line along [0 0 1] and $\phi = 0$ [1 0 0] and $\phi = 3\pi/2$ [0 1 0] directions as shown by red dotted line. The insets show all the three directions two in plane and one out of plane.

Thus, the MAE has strong dependency on polar angle and independent on the azimuthal angle. The insets in Figure 4 show the direction of spin rotation along all the three axis in the azimuthal plane.

### 3.3 Electronic Properties

We explore the electronic properties of the kagome $V_2O_3$ ML for the important applications in the electronic devices. To investigate the electronic properties, we use the GGA-PBE calculated band structure as shown in Figure 5. The red lines show the majority spin bands while the blue lines depict the minority spin bands. Figure 5 (a) shows that the kagome $V_2O_3$ ML has a large direct spin band gap of 4.38 eV in the minority spin bands located at the $\Gamma$ point, while the majority spin band shows metallic nature, which confirms the half-metallic nature of the kagome $V_2O_3$ ML.

Similarly, for band structure of kagome $V_2O_3$ ML, we also use on-site Hubbard type interaction (GGA+U) with U = 3.68 eV [50] for V 3d orbital. As the PBE calculated band structure is half-metallic while the GGA+U band structure transform from half metallicity to semiconductor nature although with almost zero band gap. Besides, we also explore the orbital projected density of states (PDOS) as shown in Figure 5 (c), which depicts that states near the Fermi level are originates from $d_{xy}$ and $d_{xz}$ orbitals. Moreover, dxy, $dx^2$ and $dz^2$ are far away from the Fermi level where, dxy and $dx^2$ orbitals are having the same energy states. Thus only the dxy and dxz orbitals are close to the Fermi level in the majority spin orientation. The minority spin states are far away from the Fermi level.

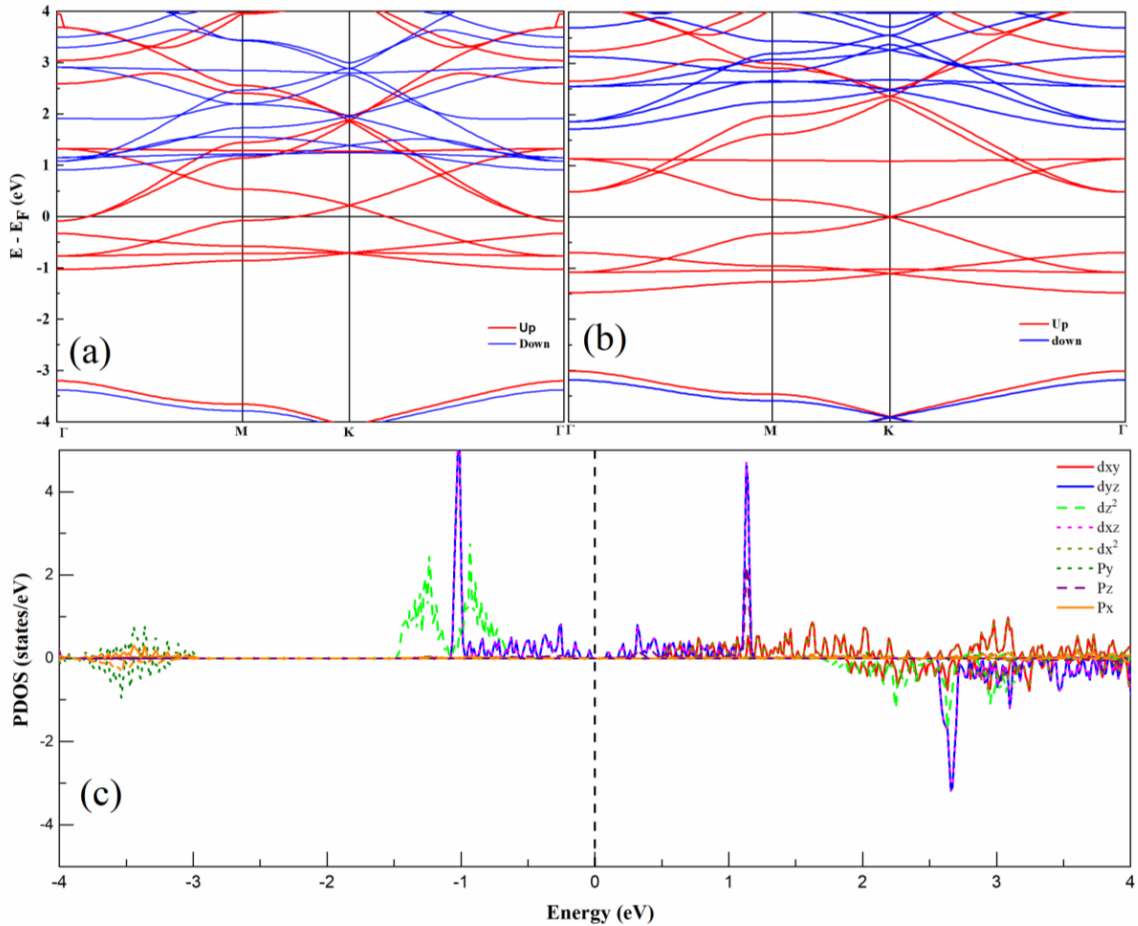

**Figure 5.** Band Structure of $V_2O_3$ ML calculated (a) with GGA-PBE, (b) GGA+U, and (c) projected density of states (PDOS) of V-d orbital and O P orbital of atom.

Further, due to the kagome lattice structure of $V_2O_3$ ML, we found two Dirac points ($D_1$, $D_2$) and two flat bands ($F_1$, $F_2$) in the GGA-PBE and GGA+U calculated band structures. Therefore, to further examine this feature we also investigated the GGA+U calculated projected band structure for $V_2O_3$ ML onto the d orbitals of V and P orbitals of the O atom as shown in Figure 6 (a)-(d) respectively. Interestingly, the Dirac point below the Fermi level is in contact with the two flat band at the Γ high symmetry point as shown by the red circles above and below the Fermi level. Now the question arises about the origin of these Dirac points. Therefore, to answer this question we explore the projected band structure as shown in Fig. 6 (a).

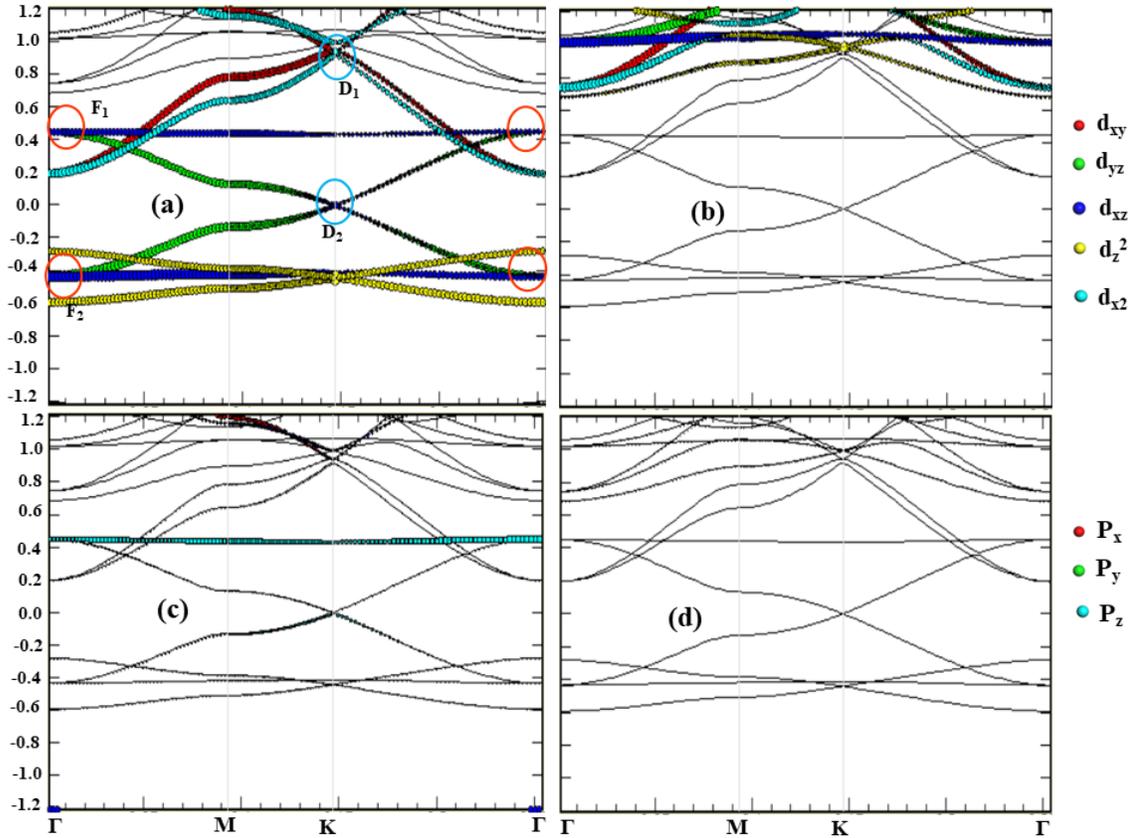

**Figure 6.** Projected Band Structure of $V_2O_3$ ML of V 3d – resoled contribution to the (a) majority and (b) minority spin and O p-resolved contribution to the (c) majority and (d) minority spin respectively.

$D_1$ is at 0.4 eV from the Fermi level and mainly originated from the hybridization of linearly dispersed bands of dxy and $dx^2$ orbitals of V atom, where dxy is contributed in the in-plane MAE while the $dx^2$ is strongly contributed in the out of plane MAE. Similarly, $D_2$ below the Fermi level is at 0.44 eV and mainly originated from dyz and dyz orbitals of V atom as shown in Fig. 3 (a). Interestingly, both these orbitals are contributed in out of plane MAE and originating from the flat bands. Generally, in transitions metal based materials, the d orbitals are dominant at the low energy physics[51] , similarly in $V_2O_3$ ML the V *d*-orbital is dominant. In flat bands, the kinetic energy of electrons in out of planes orbitals are blocked and further these orbitals are responsible for the ferromagnetism in kagome $V_2O_3$ ML. Further, there are few materials in which the flat bands are exists such as kagome lattices, twisted bilayer graphene, and heavy-fermion compounds[51]. Therefore, these out of plane d orbitals causes the ferromagnetism of the kagome $V_2O_3$ ML. Interestingly, our finding maybe a new addition to the kagome group of oxides materials.

4. **Curie Temperature**

Next, we have to discuss the Curie temperature (Tc) of the $V_2O_3$ ML. We use two different approaches for calculating Tc, the Mean Field Theory (MFT) approach and the Metropolis Monte Carlo (MC) simulation coded in the VAMPIRE software package.[52] Based on Mean-Field Theory (MFT) and the energy difference ΔE ¼ $E_{AFM}$ $E_{FM}$, the Tc could be estimated using the formula bellow

$$T_C^{MFA} = \frac{2\Delta E}{3nK_B} \quad (5)$$

, where kB is Boltzman's constant and m is the number of magnetic ions. For ΔE = 409 meV, the estimated Tc is 1579 K. Further, we also use the same approach of MFT for $CrI_3$ monolayer, and interestingly we found that the Tc of the $CrI_3$ monolayer is about 80 K. Thus, it is confirmed that the MFT overestimates the Tc, therefore, we calculate the Tc using VAMPARE software for both the $CrI_3$ and $V_2O_3$ monolayers. Interestingly, the Tc of $CrI_3$ monolayer 46, is in accordance to the experimentally reported value of the Curie temperature. Therefore, our calculated value of Tc for $V_2O_3$ ML is also expected to be in accordance with the actual value if experimentally estimated.

We performed the temperature dependent magnetization curve using the Metropolis Monte Carlo (MC) simulation coded in the VAMPIRE software package [52] and then we extracted the Curie temperature ($T_c$). Here, we used the classical spin Heisenberg model which can be written

$$H = -\sum_{i,j} J m_i \cdot m_j \quad (6)$$

Where J is the exchange coupling energy parameter of the first nearest neighbor atom and $m_i$, $m_j$ shows the magnetic moment of neighboring atoms related to each exchange interaction. Then, the $J_i = \frac{\Delta E}{Nm^2}$ where N and m are the total number of magnetic atoms per unit cell and average magnetic moment per magnetic atom in the $V_2O_3$ ML unit cell. For MC simulation, we considered a large enough supercell of 50 x 50x1. Note that the magnetization curve was fitted by using the Curie-Bloch equation in the classical limit as given below

$$m(T) = \left[1 - \frac{T}{T_C}\right]^\beta \quad (7)$$

We fitted the temperature dependent magnetization curve by using the critical exponent of 0.48. Thus, the $T_c$ of the $V_2O_3$ ML is 640 K as shown in Figure 7. Initially, the magnetic moment retains high spin state in the low temperature range and then finally drops to near zero at the critical temperature. The estimated Curie temperature of the pristine $CrI_3$ is 46 K and this agrees with the experimentally reported value of 45 K [53]. This also shows the validity of our estimated results of Curie temperature.

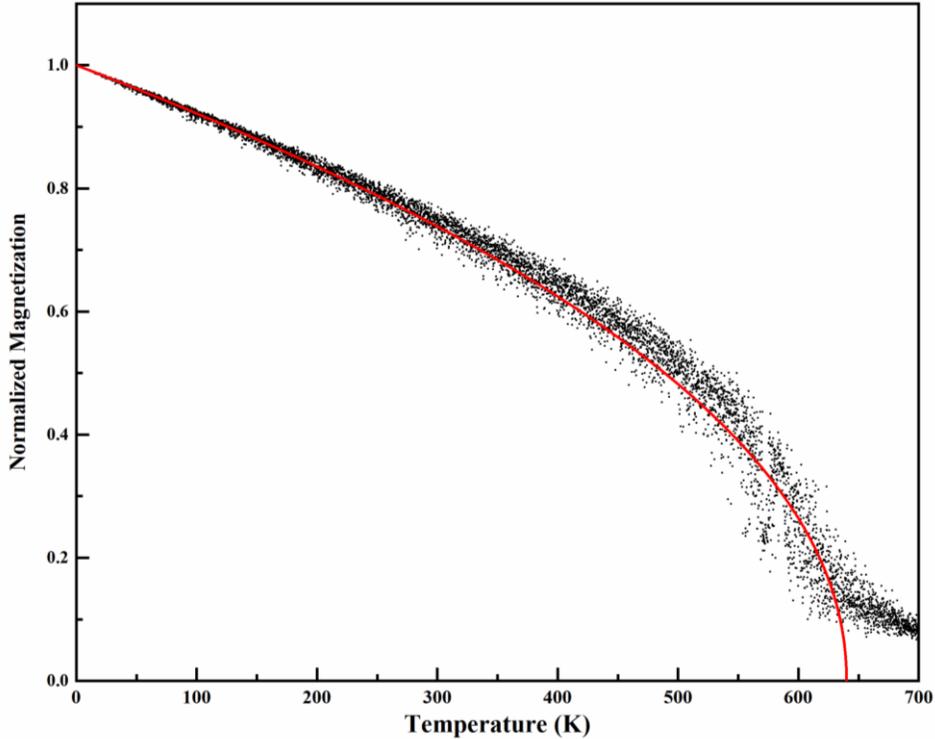

**Figure 7.** Curie temperature of the $V_2O_3$ ML, the black dotted curve shows the temperature dependent magnetization curve, while the red line shows the fitted result with Eq. (7).

## 5. Conclusions

In summary, we use the first principles calculations to explore the kagome $V_2O_3$ ML. Further, to investigate the ising ferromagnetism and half-metallicity character of the $V_2O_3$ ML, we use different functional and found different behavior towards the functional. For example, the PBE results shows half-metallicity while the on-site Hubbard type interaction for V 3d orbital gives a semiconductor band structure of zero band gap. Both GGA Perdew-Burke-Ernzerhof (PBE), and GGA+U gives us a strong ferromagnetic ground state with a constant magnetic moment of $2\mu_B$ per V atom. For calculating the magnetic anisotropy energy (MAE), we apply the spin-orbit coupling (SOC) along [1 0 0], [1 0 1], and [0 0 1] directions and obtained ~0.155 meV perpendicular magnetic anisotropy. Further, through the SOC analysis, we found that both the V atoms are equally and prominently contributed to the MAE while the O atoms contribution is almost negligible. We also analyzed the spin orientation dependence on the orbital. Besides, the Curie temperature calculated by both the Mean Field Theory and Metropolis Monte Carlo (MC) simulations and the it is further confirmed that the MFT results are overestimated as compared to the Metropolis Monte Carlo (MC) simulations. This robust and large value of the Curie temperature provide a feasible way to study the $V_2O_3$ ML at room in the experimental work.

**Author Contributions:**

F.S conceived the idea, did the DFT calculations, and plot the results. All the authors carefully check the results and help throughout the writing and editing the manuscript.

**Acknowledgments:**

This work is supported by the National Natural Science Foundation of China (Grant No. 52006057), the Fundamental Research Funds for the Central Universities (Grant Nos. 531119200237 and 541109010001), the Natural Science Foundation of Chongqing, China (No. CSTB2022NSCQ-MSX0332), and the State Key Laboratory of Advanced Design and Manufacturing for Vehicle Body at Hunan University (Grant No. 52175013). Z.Q. is supported by the National Natural Science Foundation of China (Grant No.12274374, 11904324). The numerical calculations in this paper have been done on the supercomputing system of the E.T. Cluster and the National Supercomputing Center in Changsha.

**Conflicts of Interest:**

The authors declare no conflict of interest.